\documentclass[twocolumn,english,aps,prl,10pt]{revtex4-1}
\usepackage{babel}
\usepackage{amsmath}
\usepackage{graphicx}
\usepackage{amssymb}
\usepackage{hyperref}
\usepackage{txfonts}

\hypersetup{
    colorlinks,
    linkcolor=blue,
    citecolor=blue,
    urlcolor=blue,
    linktocpage=true
}

\newcommand{\nio}{Na$_2$IrO$_3$}
\newcommand{\lio}{Li$_2$IrO$_3$}

\newcommand{\rucl}{$\alpha$-RuCl$_3$}

\newcommand{\Ztwo}{\mathbb{Z}_2}

\newcommand{\spiral}{3D spiral}
\newcommand{\AFb}{$\mathrm{AF}_{abc}$}

\graphicspath{{./}{./plots/}}

\begin{document}

\title{
Heisenberg-Kitaev Models on Hyperhoneycomb and Stripyhoneycomb Lattices: \\
3D--2D Equivalence of Ordered States and Phase Diagrams}

\author{Wilhelm G. F. Kr\"uger}
\author{Matthias Vojta}
\author{Lukas Janssen}
\affiliation{Institut f\"ur Theoretische Physik and W\"urzburg-Dresden Cluster of Excellence ct.qmat, Technische Universit\"at Dresden,
01062 Dresden, Germany}

\date{\today}

%%%%%%%%%%%%%%%%%%%%%%%%%%%%%%%%%%%%%%%%%%%%%%%%%%%%%%%%%%%%%%%%%%%%%%%

\begin{abstract}
We discuss magnetically ordered states, arising in Heisenberg-Kitaev and related spin models, on three-dimensional (3D) harmonic honeycomb lattices. For large classes of ordered states, we show that they can be mapped onto two-dimensional (2D) counterparts on the honeycomb lattice, with the classical energetics being identical in the 2D and 3D cases.
As an example, we determine the phase diagram of the classical nearest-neighbor Heisenberg-Kitaev model on the hyperhoneycomb lattice in a magnetic field: This displays rich and complex behavior akin to its 2D counterpart, with most phases and phase boundaries coinciding exactly.
To make contact with the physics of the hyperhoneycomb iridate $\beta$-\lio, we also include a symmetric off-diagonal $\Gamma$ interaction, discuss its 3D--2D mapping, and determine the relevant phase diagrams.
In particular, we demonstrate explicitly the adiabatic equivalence of the spiral magnetic orders in $\alpha$- and $\beta$-\lio.
%
%We also discover phases of the 3D models which evade the mapping, i.e., are of genuine 3D character.
%
Our results pave the way to a systematic common understanding of 2D and 3D Kitaev materials.
\end{abstract}

\date{\today}

\maketitle

%%%%%%%%%%%%%%%%%%%%%%%%%%%%%%%%%%%%%%%%%%%%%%%%%%%%%%%%%%%%%%%%%%%%%%%

In studies of quantum magnetism, materials with strong spin-orbit coupling have moved to center stage, as they promise to realize novel phases beyond those known for spin-symmetric Heisenberg models \cite{jackeli2009, nussinov2015, rau2016}. A paradigmatic example for non-trivial effects of spin-anisotropic interactions is Kitaev's celebrated honeycomb-lattice spin model, being exactly solvable and realizing a $\Ztwo$ spin liquid of Majorana fermions~\cite{kitaev2006}.
In the search for materials realizations, compounds with $4d$ and $5d$ transition-metal ions arranged on layered honeycomb lattices have been proposed \cite{jackeli2009}, such as \nio, $\alpha$-\lio, and \rucl\ \cite{singh2012, choi2012, sears2015, williams2016, winter2017b}. In these materials, the combined effect of spin-orbit coupling, Coulomb interaction, and exchange geometry generates $J_{{\rm eff}}=1/2$ moments subject to a combination of exchange interactions, the most important ones being Kitaev, Heisenberg, and symmetric off-diagonal \cite{shitade2009, chaloupka2010, rau2014a}.
These two-dimensional (2D) Kitaev materials have generated tremendous interest, and the relevant extended Kitaev models have been shown to host both spin-liquid and symmetry-broken phases \cite{chaloupka2013, price2012}.

Beyond two dimensions, it has been shown that the Kitaev model can also be exactly solved on particular three-dimensional (3D) lattices with three-fold coordination \cite{kimchi2014b, hermanns2016}, and the materials $\beta$-\lio\ \cite{biffin2014b, takayama2015} and $\gamma$-\lio\ \cite{modic2014, biffin2014c} were found to realize two of these lattices, the hyperhoneycomb and stripyhoneycomb lattices. In fact, these two are part of an infinite family of 3D lattices whose limiting case is the 2D honeycomb lattice---the harmonic-honeycomb series \cite{modic2014, kimchi2014b, kimchi2015}.
While \nio\ and \rucl\ display collinear zigzag order \cite{choi2012, ye2012, sears2015} at low temperatures, the magnetic states of the \lio\ polytypes involve non-collinear spin spirals. 
In an applied magnetic field, the spiral order in $\beta$-\lio\ is rapidly suppressed, giving way to a zigzag state analogous to those of the planar honeycomb materials \cite{ruiz2017, majumder2019}.
While a number of concrete studies of either 2D or 3D models have appeared, a common systematic understanding is lacking.

In this Letter, we consider magnetically ordered states on the 3D hyperhoneycomb and other harmonic honeycomb lattices and establish a remarkable correspondence to states on the 2D honeycomb lattice: For large classes of states, we demonstrate a precise mapping between 3D and 2D, with the ground-state and magnon-mode energies, as well as other momentum-resolved observables, such as the dynamic spin structure factor, being identical in the semiclassical limit.
As an example, we determine the phase diagram of the classical nearest-neighbor Heisenberg-Kitaev model on the hyperhoneycomb lattice in a magnetic field: We show that this is \emph{almost} identical to that of the same model on the honeycomb lattice \cite{janssen2016}, with the exception of a single field-induced phase that is of genuine 3D character and thus escapes the 3D--2D mapping.
In order to model $\beta$-\lio, we include symmetric off-diagonal $\Gamma$ interactions on the hyperhoneycomb lattice, which can induce incommensurate spiral phases, and we discuss their exact 3D--2D mapping.
Finally, we consider effects beyond the classical limit and compare quantum corrections to the magnetization between the 3D and 2D cases.

%%%%%%%%%%%%%%%%%%%%%%%%%%%%%%%%%%%%%%%%%%%%%%%%%%%%%%%%%%%%%%%%%%%%%%%

\begin{figure}
\includegraphics[width=\linewidth]{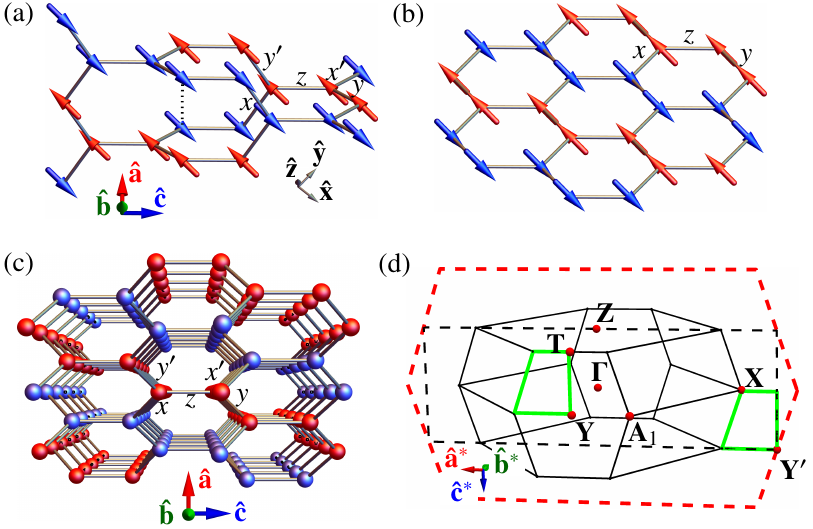}
\caption{
(a) Hyperhoneycomb lattice with magnetic skew-zigzag state and
(b) honeycomb lattice with zigzag state; both are equivalent as explained in the text.
(c) Hyperhoneycomb lattice viewed from the crystallographic $\mathbf{\hat b}$ direction, illustrating its projection onto an elongated honeycomb lattice. Colored balls indicate spin directions.
(d) Brillouin zones of the hyperhoneycomb lattice (black) and the elongated honeycomb lattice (red dashed). 3D states with ordering wavevectors in the $ac$ plane can be transformed into equivalent 2D states on the honeycomb lattice. The quarters of the front hexagon (such as the green quadrangle) can be shifted with reciprocal lattice vectors to the $ac$ plane and form, together with the $ac$-plane cut of the first Brillouin zone, a rectangle (black dashed). The latter becomes the 2D Brillouin zone of the elongated honeycomb lattice if a four-site unit cell is chosen. All high-symmetry points shown are quasi-2D.
}
\label{fig:map}
\end{figure}

\paragraph{Mapping from 3D to 2D: General.}
The hyperhoneycomb lattice is a tricoordinated 3D lattice that can be classified as face-centered orthorhombic with a four-site atomic unit cell \cite{lee2014a, hermanns2014}.
In Fig.~\ref{fig:map}(a), we show the spin configuration of a so-called skew-zigzag antiferromagnetic state on this lattice, while Fig.~\ref{fig:map}(b) displays a corresponding zigzag state on the 2D honeycomb lattice. These two states are equivalent in the following sense:
Projecting the hyperhoneycomb lattice onto its $ac$ plane results in an elongated honeycomb lattice, Fig.~\ref{fig:map}(c), and this projection transforms the skew-zigzag state from panel (a) into the zigzag state of panel (b). Importantly, their classical-limit energies are identical, because in both cases each site faces neighbors with identical alignment.
This projective equivalence, which is the central result of this Letter, applies to all 3D ordered states where sites separated by the lattice vector $\mathbf b$ [dashed line in Fig.~\ref{fig:map}(a)] are magnetically equivalent. As we will see below, this includes large classes of 3D magnetic states, which we will dub ``quasi-2D''.

This 3D--2D equivalence holds despite the fact that the symmetry groups on the two lattices are obviously different. In particular, the number of sites in the primitive unit cell on the hyperhoneycomb lattice is four, and thus twice its value on the honeycomb lattice. It is thus possible to construct states (e.g., skew-zigzag states) that do not break the translation symmetry on the hyperhoneycomb lattice, although their 2D projections break the honeycomb translation symmetry.

Further insight is gained in reciprocal space. All high-symmetry points displayed in Fig.~\ref{fig:map}(d) have a vanishing component along the direction of the reciprocal lattice vector $\mathbf{b}^*$ (up to reciprocal-lattice translations). States with ordering wavevectors at these high-symmetry points thus exhibit no modulation along the $\mathbf{b}$ axis in real space and are quasi-2D: Their projection onto the $ac$ plane yields states on the honeycomb lattice with the \emph{exact same} classical energy.
This applies to all ordered phases in the nearest-neighbor Heisenberg-Kitaev (HK) model in zero field.
Upon the inclusion of other symmetry-allowed interactions, as well as in a magnetic field, Kitaev systems also stabilize multi-$\mathbf Q$ and incommensurate states. We will show that even such more exotic states, including the counterrotating spiral states that are realized in the different \lio\ polytypes, are quasi-2D.
Moreover, the 3D--2D mapping discussed here for the hyperhoneycomb lattice can be extended to the full harmonic honeycomb series, for details see the supplemental material (SM) \cite{suppl}.
%
%%%BIBITEMSTYLER \cite{luttinger1946, lyons1960, chaloupka2015}
%
Importantly, and in contrast to other families of lattices in different dimensions, all members of the harmonic honeycomb series have the same coordination number, which is a prerequisite for the 3D--2D equivalence to hold.

%%%%%%%%%%%%%%%%%%%%%%%%%%%%%%%%%%%%%%%%%%%%%%%%%%%%%%%%%%%%%%%%%%%%%%%

\paragraph{Heisenberg-Kitaev model in a magnetic field.}
To illustrate the power of the advertised mapping, we consider the spin-$S$ HK Hamiltonian \cite{chaloupka2010, chaloupka2013} in a uniform magnetic field $\mathbf h$,
\begin{equation}
	\mathcal{H}_\text{HK} =
	J \sum_{\left\langle ij\right\rangle} \mathbf{S}_{i}\cdot\mathbf{S}_{j}
	+ K \sum_{\left\langle ij\right\rangle_{\gamma}} S_{i}^{\gamma}S_{j}^{\gamma}
	- \mathbf h \cdot \sum_{i} \mathbf S_i,
\label{hk}
\end{equation}
where $\gamma \in \{x, y, z\}$ labels the three different types of bonds on the lattice. 
The 2D model on the honeycomb lattice has been studied intensely, see Refs.~\onlinecite{rau2016, trebst2017, hermanns2018, janssen2019} for reviews.
%
%%%BIBITEMSTYLER \cite{rau2016, trebst2017, hermanns2018, janssen2019}
%
For non-zero field, the classical phase diagram (i.e., for $S\to\infty$) of this and related models has been determined \cite{janssen2016, chern2017, janssen2017, chern2019}, and the $S=1/2$ case has also been studied \cite{jiang2011, gohlke2018, zhu2018, ronquillo2018, jiang2018, zou2018, hickey2019, patel2018,  nasu2018, liang2018}.
The 3D model on the hyperhoneycomb lattice in zero field has been considered in Ref.~\onlinecite{lee2014a}.
%
%%%BIBITEMSTYLER \cite{lee2014a, lee2014b}

%%%%%%%%%%%%%%%%%%%%%%%%%%%%%%%%%%%%%%%%%%%%%%%%%%%%%%%%%%%%%%%%%%%%%%%
\begin{figure*}
\includegraphics[width=\linewidth]{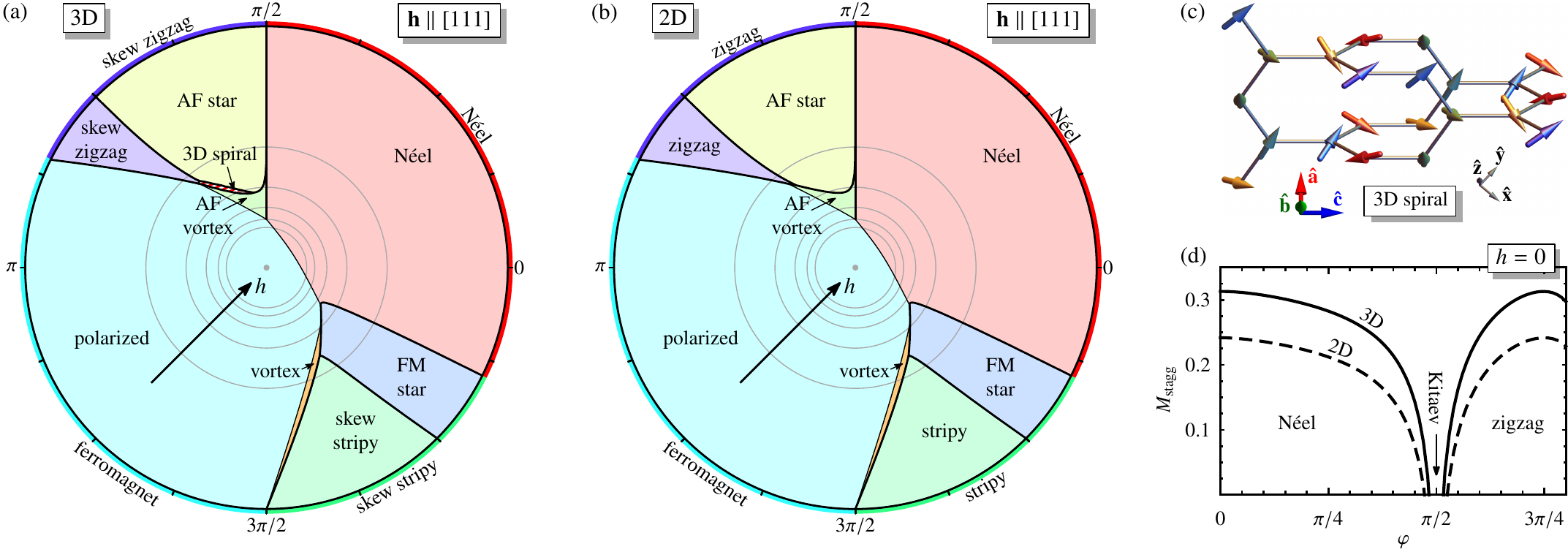}
\caption{
Phase diagram of the classical HK model in a magnetic field $\mathbf h$ along the $[111] \propto (\mathbf{\hat x}+ \mathbf{\hat y}+ \mathbf{\hat z})/\sqrt{3}$ direction for $T\to0$, with $J=A \cos \varphi$, $K = 2 A \sin \varphi$ \cite{circlenote}, and the radial direction representing the field strength $h$, with $h/(AS) = 1, 2, 3, 4, 5$ from outer to inner gray circles, on the
(a) hyperhoneycomb lattice and
(b) honeycomb lattice.
The latter agrees with the previous analysis~\cite{janssen2016, chern2017}, with the exception of small regions for which the true ground state has incommensurate ordering wavevectors.
Thick (thin) black lines denote first-order (second-order) phase transitions.
Except for the \spiral\ phase, for which a representative spin configuration is shown in (c), all phases of the hyperhoneycomb lattice have a 2D analogue with exactly the same ground-state energy.
(d) Order parameters in N\'eel and zigzag phases from linear spin-wave theory for $S=1/2$ and zero field, comparing the hyperhoneycomb (solid) and honeycomb (dashed) lattices.
}
\label{fig:hkpd}
\end{figure*}
%%%%%%%%%%%%%%%%%%%%%%%%%%%%%%%%%%%%%%%%%%%%%%%%%%%%%%%%%%%%%%%%%%%%%%%

We have determined the phase diagram of the classical HK model in a magnetic field using a combination of high-field spin-wave theory and classical energy minimization. On the hyperhoneycomb lattice, the number of possible geometries of the magnetic unit cell drastically increases with its size. For reasons of numerical feasibility and consistency, we have restricted the numerical energy minimization in both the 2D and 3D cases to states with up to 12 sites in the magnetic unit cell, but have included all possible unit-cell geometries \cite{suppl}. On the honeycomb lattice, our findings are consistent with the previous analyses \cite{janssen2016, ICnote}.

The result, comparing the hyperhoneycomb and honeycomb cases for a field along the $[111] \propto (\mathbf{\hat x}+\mathbf{\hat y}+\mathbf{\hat z})/\sqrt{3}$ direction, is displayed in Fig.~\ref{fig:hkpd}. 
The various phases are characterized in the SM~\cite{suppl}.
Remarkably, both phase diagrams agree \emph{quantitatively}, with the exception of the \spiral\ phase, which appears only in the 3D case of panel (a). Inspecting the individual phases, we see that all phases except the \spiral\ have ordering wavevectors located in the $ac$ plane, such that 3D--2D mapping applies, whereas the \spiral\ phase has $\mathbf Q= \frac{1}{3}\mathbf{b}^*$, evading the mapping. The latter is hence a genuine 3D phase, consisting of spirals along the $\mathbf b$ direction in a 12-site magnetic unit cell, see Fig.~\ref{fig:hkpd}(c) and Ref.\ \onlinecite{suppl}.

We note that early work on the hyperhoneycomb-HK model in a magnetic field~\cite{lee2014b} missed the nontrivial field-induced phases found here. We have explicitly checked that our novel intermediate phases have lower energies than the canted skew-zigzag and skew-stripy states suggested in Ref.~\onlinecite{lee2014b}.
%
%%%BIBITEMSTYLER \cite{lee2014b}

%%%%%%%%%%%%%%%%%%%%%%%%%%%%%%%%%%%%%%%%%%%%%%%%%%%%%%%%%%%%%%%%%%%%%%%

\paragraph{$\Gamma$ and other interactions.}
For actual Kitaev materials, it has been shown that, in addition to nearest-neighbor Kitaev and Heisenberg interactions, also symmetric off-diagonal interactions, commonly dubbed $\Gamma$ interactions, are important \cite{rau2014a, winter2016, janssen2017, lee2015, lee2016, ducatman2018, rousochatzakis2018}. To model $\beta$-\lio, we hence consider the Hamiltonian
\begin{equation} \label{hkg}
	\mathcal{H}_\mathrm{HK\pm\Gamma}=
	\sum_{\langle ij\rangle_\gamma}\left[
	J \mathbf{S}_i \cdot\mathbf{S}_j
	+ KS^\gamma_iS^\gamma_j
	\pm\Gamma\left(S^\alpha_iS^\beta_j+S^\beta_iS^\alpha_j\right)\right],
\end{equation}
where $\alpha$ and $\beta$ label the two remaining directions on a $\gamma$ bond. On the hyperhoneycomb lattice, there are two inequivalent types of pairwise parallel $x$ ($y$) bonds, denoted as $x$ and $x'$ ($y$ and $y'$), respectively, while all $z$ bonds are equivalent; see Fig.~\ref{fig:map}(a). We take the upper (lower) sign in front of the $\Gamma$ interaction on $x$, $y$, and $z$ ($x'$ and $y'$) bonds; this choice can be justified microscopically~\cite{lee2015}.
Applying the 3D--2D mapping to this model, which we dub HK$\pm\Gamma$ model, we see that it corresponds to an unusual 2D model. Compared to the Heisenberg-Kitaev-$\Gamma$ (HK$\Gamma$) model typically considered for 2D Kitaev materials, this has a supermodulation in the $\Gamma$ interaction.
Remarkably, in the limit of small $J$, the HK$\pm\Gamma$ and HK$\Gamma$ models can be shown to be dual to each other by using a unitary transformation that rotates the spins on consecutive zigzag chains by $\pi/2$ and $3\pi/2$, respectively, about the $z$ axis~\cite{suppl}.
In the presence of sizable $\Gamma$ interactions, incommensurate states appear \cite{rau2014a, lee2015}.

%%%%%%%%%%%%%%%%%%%%%%%%%%%%%%%%%%%%%%%%%%%%%%%%%%%%%%%%%%%%%%%%%%%%%%%
\begin{figure*}
\includegraphics[width=\linewidth]{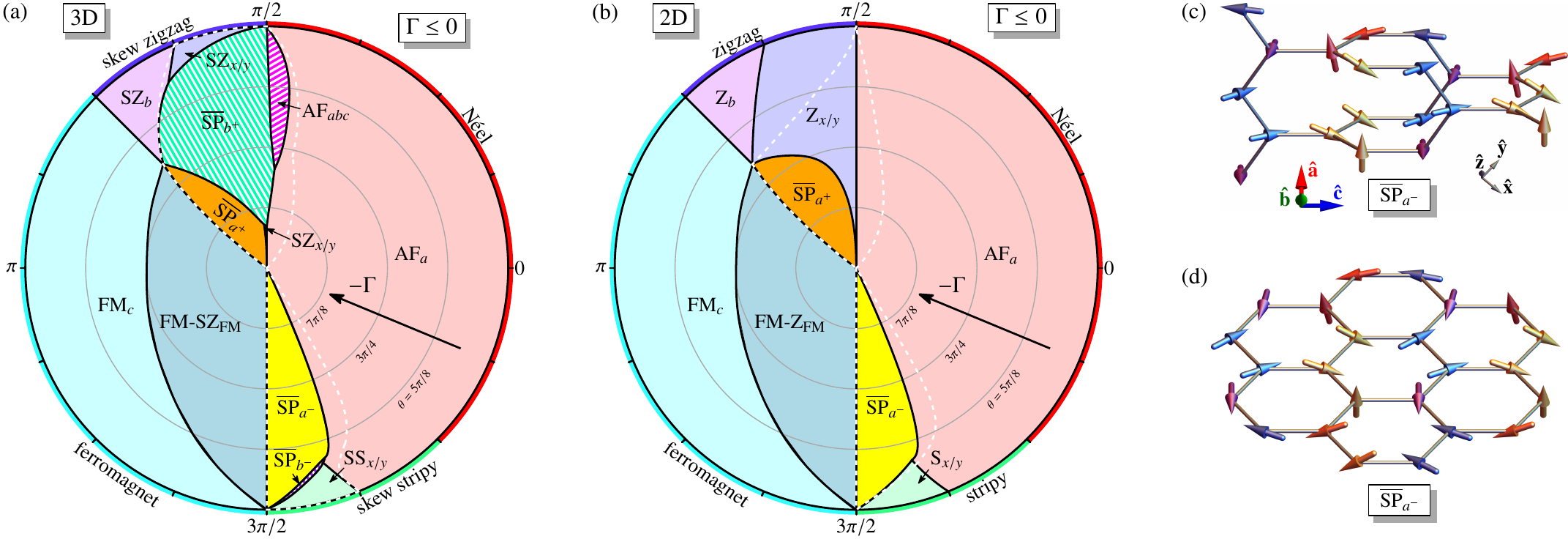}
\caption{
Phase diagram of the classical HK$\pm\Gamma$ model for $T\to0$, with $J=A \sin\theta \cos \varphi$, $K = A \sin\theta \sin \varphi$, and $\Gamma = A \cos\theta \leq 0$ \cite{circlenote} on the
(a) hyperhoneycomb lattice \cite{lee2015} and
(b) honeycomb lattice.
Hatched regions in (a) denote genuine 3D phases with ordering wavevectors outside the $ac$ plane.
The white dashed line indicates the regions in which the Luttinger-Tisza approach fails to satisfy the local length constraint and a single-$\mathbf Q$ ansatz has been employed instead; see SM~\cite{suppl}.
The 3D--2D equivalence holds also for the incommensurate spiral phase $\overline{\mathrm{SP}}_{a^-}$, relevant for $\beta$-\lio, for which a representative spin configuration on the hyperhoneycomb lattice is plotted in (c), together with its projection onto the honeycomb lattice in (d).
}
\label{fig:hkgpd}
\end{figure*}
%%%%%%%%%%%%%%%%%%%%%%%%%%%%%%%%%%%%%%%%%%%%%%%%%%%%%%%%%%%%%%%%%%%%%%%

Despite these complications, the concept of the 3D--2D mapping continues to apply. We illustrate this in Fig.~\ref{fig:hkgpd}, where we show the classical phase diagrams of the HK$\pm\Gamma$ model at zero field for the hyperhoneycomb and honeycomb lattices. These have been obtained via a combination of a Luttinger-Tisza analysis and a single-$\mathbf Q$ ansatz (as the finite-cluster minimization does not capture incommensurate states); see the SM for details~\cite{suppl}. Our result on the hyperhoneycomb lattice agrees with the previous work~\cite{lee2015}, except for a small region around the \AFb\ phase, for which the true ground state may be a multi-$\mathbf Q$ state that is beyond our ansatz.
Again, the 3D and 2D phase diagrams agree \emph{quantitatively}, with the exception of the $\overline{\mathrm{SP}}_{b^\pm}$ and \AFb\ phases, which appear only in panel (a) and are thus genuinely 3D.
This result is particularly striking for the counterrotating spiral $\overline{\mathrm{SP}}_{a^\pm}$ phases, for which the ground state is incommensurate, but with an ordering wavevector $\mathbf Q \parallel \mathbf{a}^*$, i.e., within the $ac$ plane; cf.\ Fig.~\ref{fig:hkgpd}(c,d).
The $\overline{\mathrm{SP}}_{a^-}$ phase includes the ground state realized in $\beta$-\lio\ \cite{biffin2014b, lee2015, lee2016}, such that 3D--2D mapping directly applies to this material.
For small $J$, the $\overline{\mathrm{SP}}_{a^-}$ phase becomes commensurate with ordering wavevector $\mathbf Q = \frac13 \mathbf{a}^*$~\mbox{\cite{rousochatzakis2018, ducatman2018}}. The 3D--2D equivalence, together with the above-mentioned duality between the HK$\pm\Gamma$ and HK$\Gamma$ models, maps this state to the coplanar $120^\circ$ state on the honeycomb lattice \cite{rau2014a}, which is indeed close to the observed magnetic order in the planar polytype $\alpha$-\lio\ \cite{williams2016}. In fact, the mapping explains several characteristic common features observed in the different \lio\ polytypes: (i) As a consequence of the structure of the duality transformation, the hyperhoneycomb-lattice state corresponding to the $120^\circ$ state consists of zigzag chains with coplanar spins, in agreement with the experimental findings in $\beta$-\lio~\cite{biffin2014b}. (ii) In the 120$^\circ$ phase on the honeycomb lattice, the spins on the two sublattices  rotate in opposite directions, which explains the emergence of counterrotating spirals on the hyperhoneycomb lattice. (iii) The duality between $\overline{\mathrm{SP}}_{a^-}$ and $120^\circ$ states for small $J$ furthermore reveals why the angle between every second spin of a zigzag chain is close to $120^\circ$ in both $\alpha$- and $\beta$-\lio\ \cite{williams2016, biffin2014b}.

For completeness, we note that the 3D--2D mapping can be extended to interactions beyond nearest neighbors; for details see the SM~\cite{suppl}.

\paragraph{Beyond the classical limit.}
While the advertised \emph{qualitative} 3D--2D mapping of ordered states is very general, their \emph{quantitative} energetic equivalence only applies to the classical limit, $S\to\infty$. We have therefore studied quantum effects in a systematic $1/S$ expansion using spin-wave theory.
A first remarkable insight is that the leading-order magnon spectra and the dynamic structure factors also follow the 3D--2D mapping, i.e., the magnon energies and weights on the different harmonic honeycomb lattices are \emph{identical} for 3D wavevectors belonging to the $ac$ plane; this is demonstrated explicitly in the SM~\cite{suppl}.

In Fig.~\ref{fig:hkpd}(d), we display the order parameters (i.e., staggered magnetizations) evaluated for $S=1/2$ in the N\'eel and zigzag phases of the HK model at zero field, comparing the hyperhoneycomb and honeycomb lattices.
In the hyperhoneycomb case, the quantum corrections to the classical value $M_\text{stagg} = 1/2$ are smaller, but 
%the overall shape of the order parameter as function of $\varphi = \arg(2J + \mathrm i K)$ is similar to those of the honeycomb lattice. 
the qualitative behavior of the order parameter is similar to those of the honeycomb lattice.
In particular, in linear spin-wave theory, the critical couplings at which the order parameters vanish, indicating the transition to the Kitaev spin liquid phase, roughly agree. This suggests that the Kitaev spin liquid on the hyperhoneycomb lattice \cite{mandal2009} covers a parameter range that is only slightly smaller than those of its honeycomb-lattice counterpart.
In the SM~\cite{suppl}, we also show the quantum corrections to the uniform magnetization in the high-field phase.

These results illustrate that the different phase space renders quantum fluctuations stronger in 2D compared to 3D. As a result, phase boundaries will shift and spoil the exact 3D--2D equivalence for $S < \infty$. This also means that classical phases that are destroyed by quantum fluctuations in 2D possibly survive in the 3D case.

%%%%%%%%%%%%%%%%%%%%%%%%%%%%%%%%%%%%%%%%%%%%%%%%%%%%%%%%%%%%%%%%%%%%%%%

\paragraph{Summary.}
In this paper, we have established an exact correspondence between magnetically ordered spin states on the 3D harmonic honeycomb lattices and the 2D planar honeycomb lattice. This correspondence is quantitative in the classical limit and applies to large classes of ordered states.
The condition is that the respective 3D ordering wavevector(s) lie(s) in the $ac$ plane (up to reciprocal-lattice translations), which pertains to all high-symmetry points in the Brillouin zone.
We have demonstrated this 3D--2D mapping for the hyperhoneycomb-lattice Heisenberg-Kitaev model in a magnetic field, where we found exact agreement with the 2D case, with the exception of one intermediate phase which is of genuine 3D character.

The hyperhoneycomb material $\beta$-\lio\ orders in a counterrotating-spiral ground state at low temperatures~\cite{biffin2014b, takayama2015}. 
Our 3D--2D mapping, together with a duality transformation, demonstrates that this state can be understood as an adiabatic deformation of the $120^\circ$ degree state on the honeycomb lattice, which is close to the magnetic order in $\alpha$-\lio~\cite{williams2016}.
This result establishes the equivalence of the experimentally observed spiral states in the different \lio\ polytypes. $\beta$-\lio\ exhibits a nontrivial behavior in a finite magnetic field~\cite{ruiz2017, majumder2019, ducatman2018}. The 3D--2D equivalence suggests that similarly interesting in-field effects may occur also in $\alpha$- and $\gamma$-\lio\ \cite{choi2019, modic2017}. 
Together, our work paves the way to a unified understanding of the magnetism in 3D and 2D Kitaev materials and opens novel perspectives for dimensional diversification.

%%%%%%%%%%%%%%%%%%%%%%%%%%%%%%%%%%%%%%%%%%%%%%%%%%%%%%%%%%%%%%%%%%%%%%%

\begin{acknowledgments}
We thank N.\ B.\ Perkins for enlightening discussions and E.\ C.\ Andrade for collaboration on earlier related work.
This research was supported by the DFG through SFB 1143 (project id 247310070), the W\"urzburg-Dresden Cluster of Excellence on Complexity and Topology in Quantum Matter -- \textit{ct.qmat} (EXC 2147, project id 39085490), and the Emmy Noether program (JA2306/4-1, project id 411750675).
\end{acknowledgments}

%%%%%%%%%%%%%%%%%%%%%%%%%%%%%%%%%%%%%%%%%%%%%%%%%%%%%%%%%%%%%%%%%%%%%%%

%%%SORT REFS BY USING:
%%%python2.7 LaTeX-BibitemStyler.py hkmap_prl.tex hkmap_refs.tex hkmap_refs-sort.tex
%\input{hkmap_refs.tex}


\begin{thebibliography}{99}
\bibitem{circlenote} We follow the parametrization of Heisenberg and Kitaev couplings in terms of the angle $\varphi$ as used in the literature, which differ by a factor of two for the HK and HK$\pm\Gamma$ models: 
$K/J = 2 \tan\varphi$ (HK) and $K/J = \tan\varphi$ (HK$\pm\Gamma$), respectively.
$A>0$ denotes an overall energy scale.
\end{thebibliography}
\end{document}

\title{Supplemental Material for:\\
Heisenberg-Kitaev Models on Hyperhoneycomb and Stripyhoneycomb Lattices: \\
3D--2D Equivalence of Ordered States and Phase Diagrams}

\author{Wilhelm G. F. Kr\"uger}
\author{Matthias Vojta}
\author{Lukas Janssen}
\affiliation{Institut f\"ur Theoretische Physik and W\"urzburg-Dresden Cluster of Excellence ct.qmat, Technische Universit\"at Dresden,
01062 Dresden, Germany}

\date{\today}

\maketitle
%%%%%%%%%%%%%%%%%%%%%%%%%%%%%%%%%%%%%%%%%%%%%%%%%%%%%%%%%%%%%%%%%%%%%%%

\section{3D--2D mapping: Generalizations}

In the main paper, the mapping of magnetically ordered states from 3D honeycomb-lattice variants to the planar honeycomb lattice has been explicitly shown for the hyperhoneycomb lattice; the mapping of Hamiltonian terms was restricted to nearest-neighbor interactions. Here we show that the mapping applies in fact more generally.

%%%%%%%%%%%%%%%%%%%%%%%%%%%%%%%%%%%%%%%%%%%%%%%%%%%%%%%%%%%%%%%%%%%%%%%
\begin{figure*}[!bt]
\includegraphics[width=\linewidth]{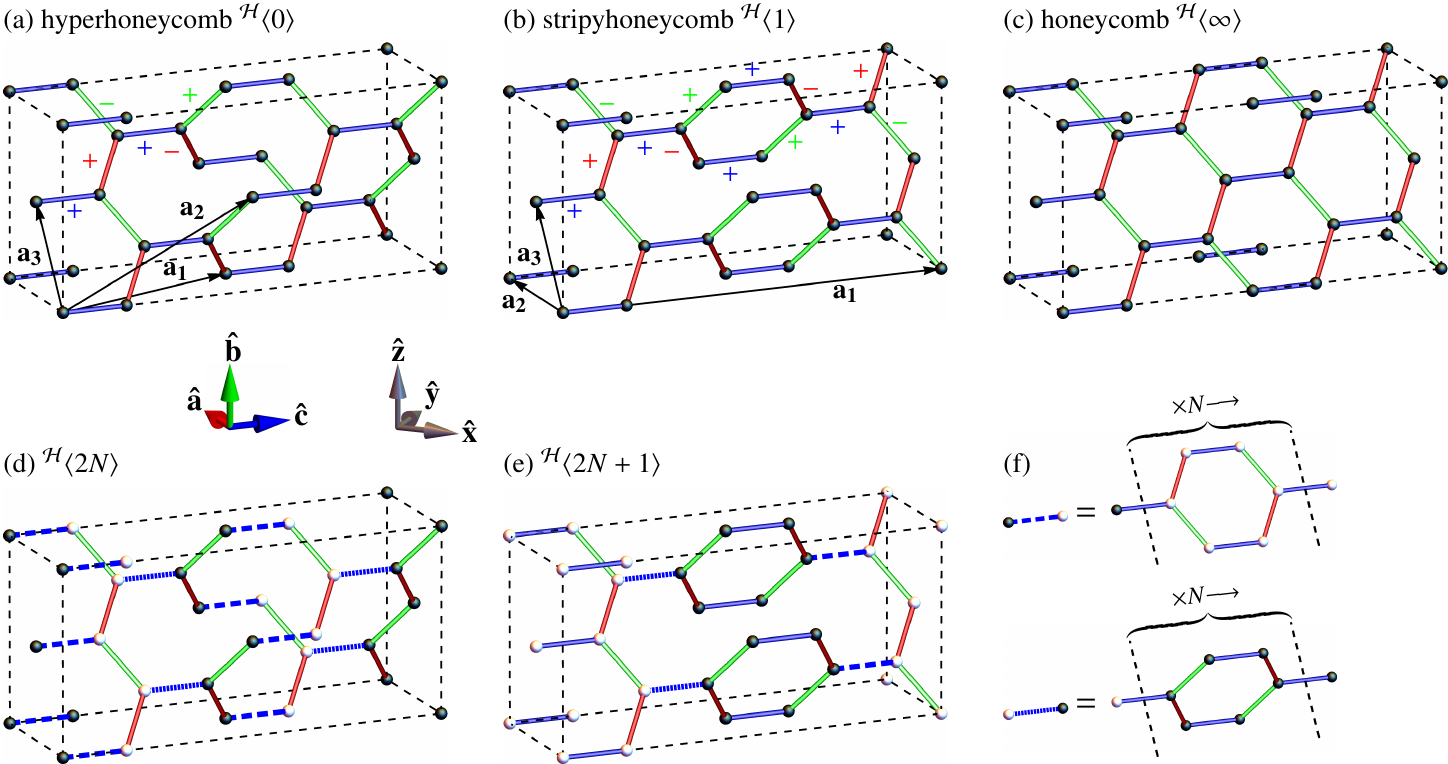}
\caption{
Harmonic honeycomb lattice series ${}^{\mathcal H}\langle N \rangle$, where $N$ refers to the number of complete honeycomb rows along the $\mathbf c$ axis \cite{kimchi2014b, modic2014}. (a)~$N=0$ (hyperhoneycomb), (b) $N=1$ (stripyhoneycomb), (c) $N \to \infty$ (honeycomb),
(d-f) building blocks for general $N$.
$x$, $y$, and $z$ bonds are marked in red, green, and blue.
The labels ``$\pm$'' refer to the signs of the $\Gamma$ interaction on the different bonds.
}
\label{fig:harmoniclatt}
\end{figure*}

%%%%%%%%%%%%%%%%%%%%%%%%%%%%%%%%%%%%%%%%%%%%%%%%%%%%%%%%%%%%%%%%%%%%%%%

\subsection{Harmonic honeycomb lattice series}

A family of so-called harmonic honeycomb lattices has been introduced in Refs.~\onlinecite{kimchi2014b, modic2014}. They consist of an arrangement of $N$ complete planar honeycomb rows along the $\mathbf c$ axis, followed by a row of $\mathbf c$-axis bonds which rotate between honeycomb planes, see Fig.~\ref{fig:harmoniclatt}. For $N=\infty$ this yields the planar honeycomb lattice, while any finite $N$ corresponds to a 3D lattice.
The case $N=0$ is commonly called hyperhoneycomb; the corresponding lattice basis vectors can be written as
$\mathbf{a}_1 = (2,4,0)$, $\mathbf{a}_2 = (3, 3, 2)$, $\mathbf{a}_3 = (-1, 1,2)$,
yielding the reciprocal basis vectors
$\mathbf{b}_1 = (\pi/3,-2\pi/3, \pi/2)$, $\mathbf{b}_2 = (-2\pi/3, \pi/3, -\pi/2)$, $\mathbf{b}_3 = (2\pi/3, -\pi/3, -\pi/2)$.
The case $N=1$ is dubbed stripyhoneycomb, with basis vectors
$\mathbf{a}_1 = (6,6,0)$, $\mathbf{a}_2 = (-2, 2, 0)$, $\mathbf{a}_3 = (-1, 1,2)$.
Parenthetically, we remark that Ref.~\onlinecite{modic2014} also discussed more general structures with more than one row of rotated bonds interspersed between planar bonds; these will not be considered here.

The 3D lattices displayed in Fig.~\ref{fig:harmoniclatt} are orthorhombic. In all cases, projecting the structure onto the $ac$ plane yields a (distorted) honeycomb lattice; this operation projects sites separated by (multiples of) the lattice vector $\mathbf b$ onto a single site. Consequently, the mapping of ordered states advertised in the main paper applies unchanged. In momentum space, the family of mappable (i.e., quasi-2D) states is characterized by ordering wavevectors with vanishing component along $\mathbf b^\ast$.

%%%%%%%%%%%%%%%%%%%%%%%%%%%%%%%%%%%%%%%%%%%%%%%%%%%%%%%%%%%%%%%%%%%%%%%
\begin{figure}
\begin{flushleft}\footnotesize (a) \hspace{.48\linewidth} (b)\hfill\end{flushleft}\vspace{-\baselineskip}
\includegraphics[scale=.28]{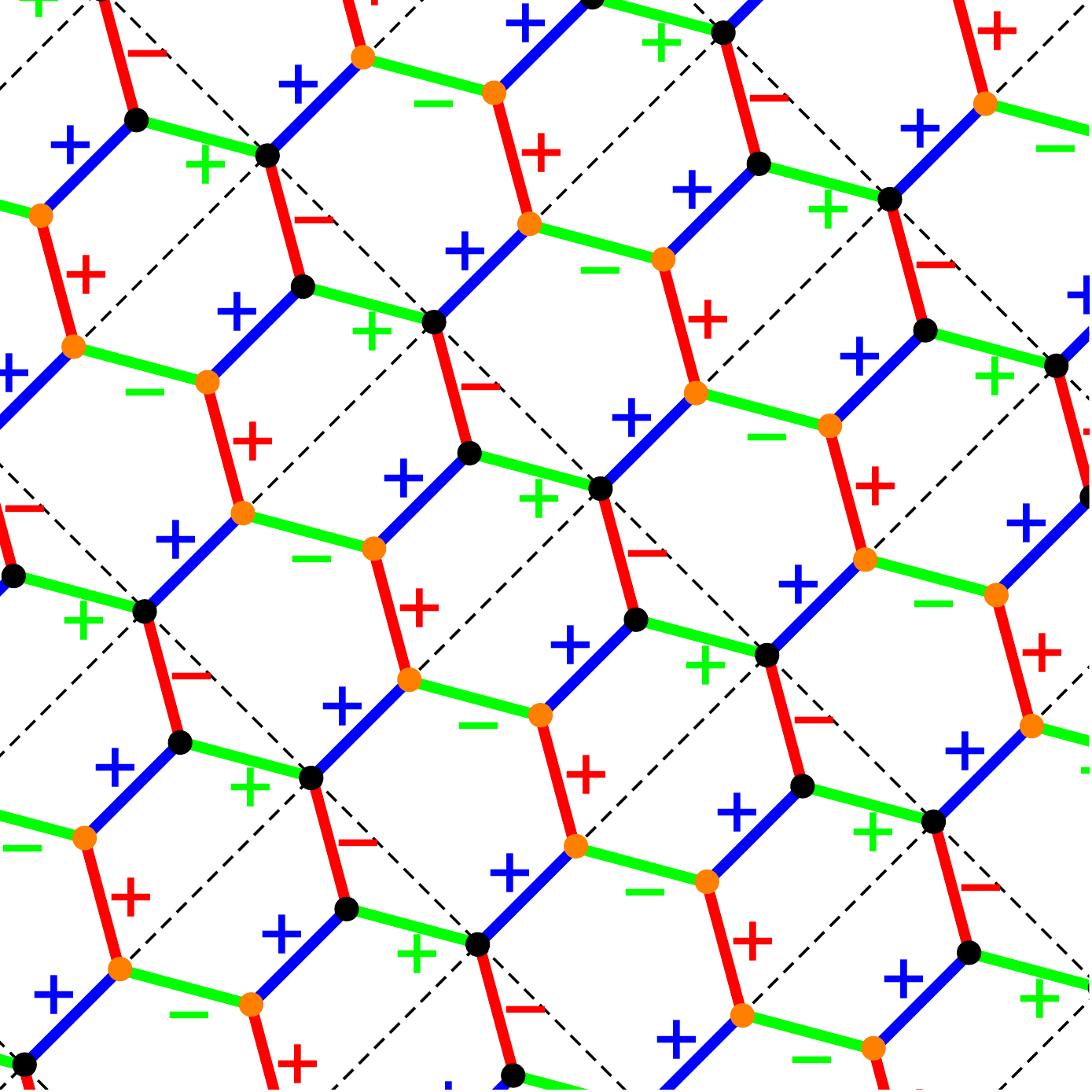}\hfill
\includegraphics[scale=.28]{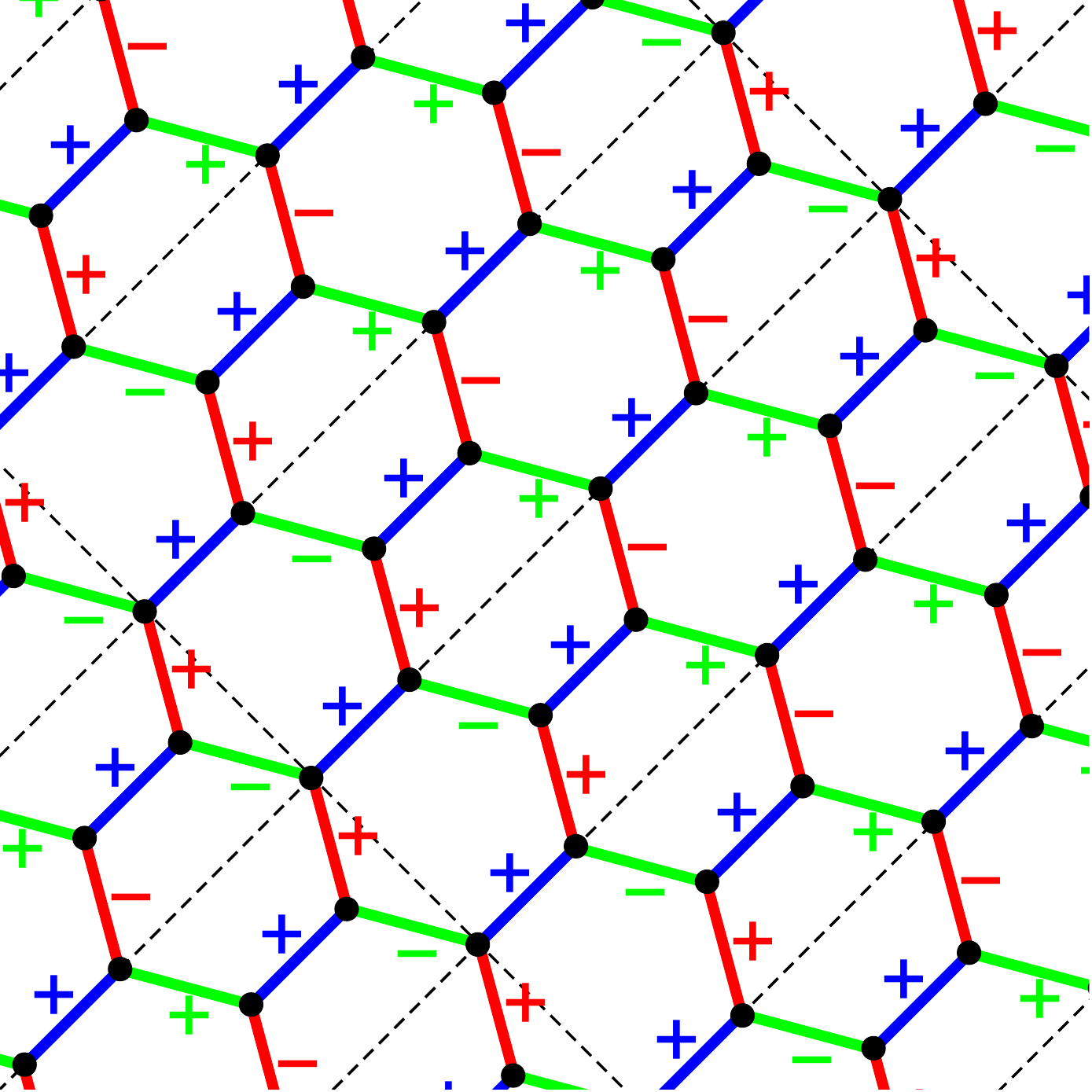}\\
\caption{
Signs of the $\Gamma$ interaction on the different bonds in the projected hyperhoneycomb-lattice (a) and stripyhoneycomb-lattice (b) models. $x$, $y$, and $z$ bonds are marked in red, green, and blue. Dashed rectangles indicate the four-site (a) and eight-site (b) primitive unit cells, respectively. The black and orange sites in (a) refer to the unitary transformation that removes the sign structure as explained in the text.
}
\label{fig:gammasign}
\end{figure}
%%%%%%%%%%%%%%%%%%%%%%%%%%%%%%%%%%%%%%%%%%%%%%%%%%%%%%%%%%%%%%%%%%%%%%%

\subsection{Symmetries and supermodulations}

The 3D--2D mapping of states is paralleled by a mapping of the underlying Hamiltonians, as shown in the main paper. Here symmetry considerations are crucial.

The 3D lattices of the harmonic honeycomb series have a crystallographic unit cell of $4+4N$ sites, to be contrasted with the two-site unit cell of the 2D honeycomb lattices. Moreover, the point-group symmetry of the (projected) 3D lattice is lower than that of the honeycomb lattice. For instance, the three types of bonds are no longer related by simple $\mathbb{Z}_3$ rotations.
Together, this implies a larger number of symmetry-allowed distinct interaction terms in 3D as compared to 2D. The 3D--2D mapping then generically leads to a Hamiltonian which breaks lattice symmetries of the honeycomb lattice.

For the hyperhoneycomb lattice of $\beta$-\lio, there are two inequivalent types of pairwise parallel $x$ ($y$) bonds, while all $z$ bonds are equivalent. The Kitaev and Heisenberg interactions are typically approximated to be of equal strength on all bonds. Consequently, this hyperhoneycomb HK model maps onto the standard 2D HK model, as used in the main text.
In contrast, the symmetric off-diagonal $\Gamma$ interaction relevant for $\beta$-\lio\ has an alternating sign structure on the $x$ and $y$ bonds.\cite{lee2015} This sign structure is shown in Fig.~\ref{fig:gammasign}, together with the corresponding sign structure on the stripyhoneycomb lattice.
The 3D--2D mapping hence produces a 2D model with a supermodulation in the $\Gamma$ interaction, dubbed HK$\pm\Gamma$ model.

\subsection{Duality transformation}

Remarkably, the alternating sign structure of the $\Gamma$ interaction in the HK$\pm\Gamma$ model, Fig.~\ref{fig:gammasign}, can be removed by a unitary transformation, if the Heisenberg coupling $J$ vanishes. This duality transformation involves a spin rotation about the $z$ axis by $\pi/2$ and $3\pi/2$ on the two families of sites shown in Fig.~\ref{fig:gammasign} in black and orange, respectively, together with a lattice reflection across a $z$ bond. On $x$ and $y$ bonds, the transformation leaves the Heisenberg and Kitaev couplings invariant while it turns all $\pm\Gamma$ couplings into $\Gamma$ couplings. By contrast, on the $z$ bonds, where the two involved sites are rotated by a different angle, the couplings transform as $(J,K,\Gamma)\to(-J,K+2J,\Gamma)$.
As a result, the K$\Gamma$ and K$\pm\Gamma$ models are thermodynamically equivalent. The 3D--2D mapping thus connects, e.g., ordered states of the hyperhoneycomb K$\pm\Gamma$ model to those of  the honeycomb K$\Gamma$ model.

\subsection{Interactions beyond nearest neighbors}

The 3D--2D mapping of Hamiltonian terms can be extended beyond the simplest nearest-neighbor exchange interactions. However, the lower symmetry of the 3D lattices complicates the discussion of further-neighbor interactions: pairs of sites with the same Euclidean distance may be symmetry-inequivalent.

\begin{figure}
\includegraphics[scale=.28]{{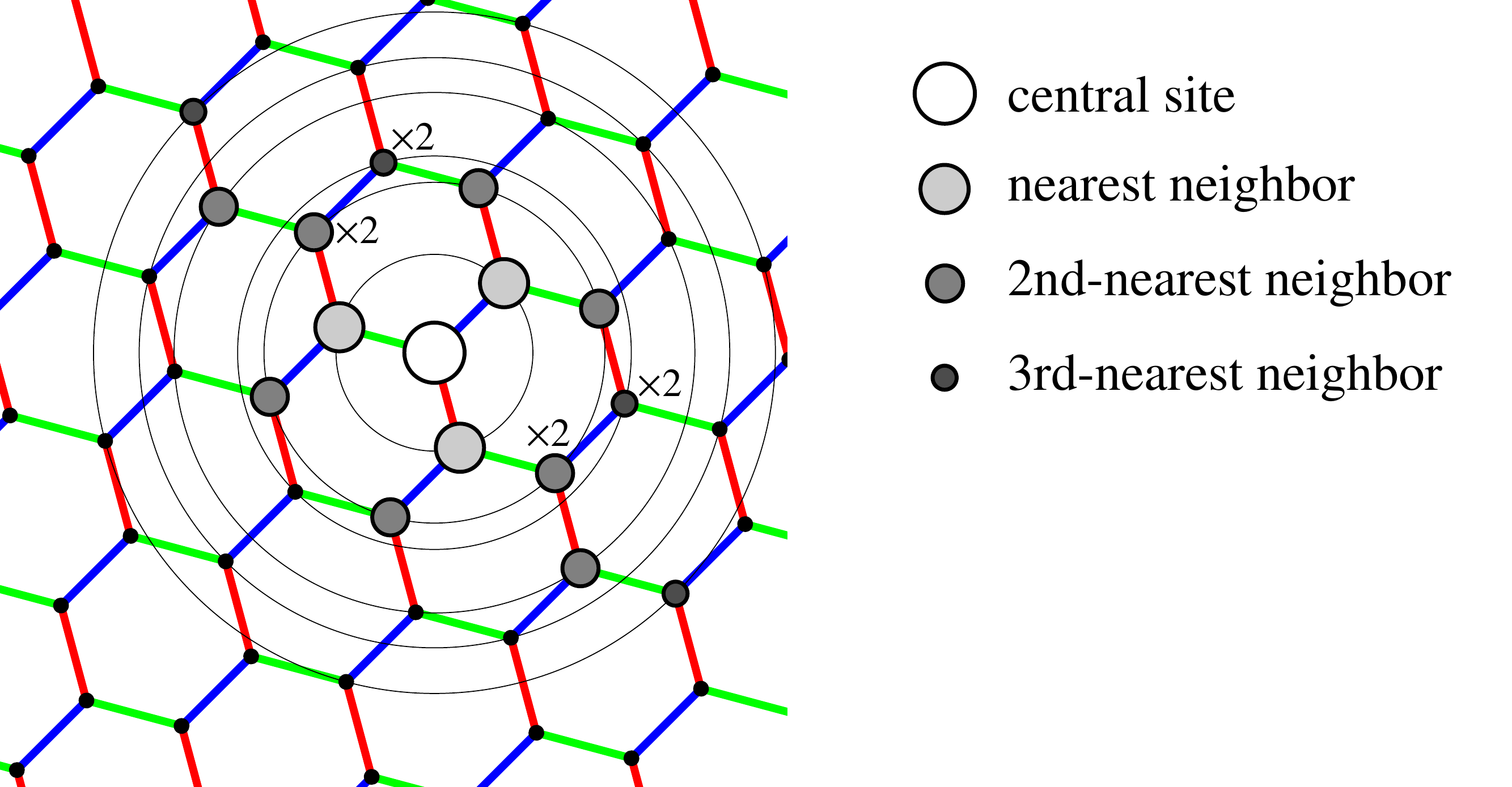}}
\caption{Projection of nearest-, second-nearest-, and third-nearest-neighbor interactions of an arbitrary central site on the hyperhoneycomb lattice. Under the projection, nearest neighbors remain nearest neighbors, while second-nearest (third-nearest) neighbors map to second- and fourth-nearest (third- and sixth-nearest) neighbors on the honeycomb lattice. ``$\times2$'' indicates cases in which two hyperhoneycomb sites map to the same honeycomb sites.
}
\label{fig:neighbors}
\end{figure}

For instance, on the hyperhoneycomb lattice, there are two types of ``second-neighbor bonds'': one where the two sites are connected by two first-neighbor bonds, and one where this connection is missing due to the twisted row of bonds. Upon projecting to the $ac$ plane, the former become second-neighbor bonds on the 2D honeycomb lattice, while the latter become fourth-neighbor bonds. Similarly, third-neighbor bonds (according to Euclidean distance) become either third-neighbor or sixth-neighbor bonds upon projection. This is illustrated in Fig.~\ref{fig:neighbors}.

Of course, the quantum chemistry underlying the actual exchange couplings (direct exchange vs.\ superexchange etc.), which in most cases also involves non-magnetic ions placed in between the magnetic ones, may discriminate symmetry-inequivalent bonds of the same Euclidean length. This physics is specific to each particular compound and requires \textit{ab-initio} consideration that are beyond the scope of this paper.

%%%%%%%%%%%%%%%%%%%%%%%%%%%%%%%%%%%%%%%%%%%%%%%%%%%%%%%%%%%%%%%%%%%%%%%
%%%%%%%%%%%%%%%%%%%%%%%%%%%%%%%%%%%%%%%%%%%%%%%%%%%%%%%%%%%%%%%%%%%%%%%

%%%%%%%%%%%%%%%%%%%%%%%%%%%%%%%%%%%%%%%%%%%%%%%%%%%%%%%%%%%%%%%%%%%%%%%

\section{Classical phases and phase diagrams}

Here we describe the methods used to determine the classical phase diagrams shown in Figs.~2 and 3 of the main paper. Due to the different nature of the expected phases we have used different methods in the two cases.

\subsection{Multi-$\mathbf Q$ phases of the HK model in a field}

For the HK model in an applied field, a large fraction of the phase diagram is covered by commensurate phases, a number of them showing multiple modulation wavevectors (multi-$\mathbf Q$ states).\cite{janssen2016, chern2017}
Hence, it is efficient to assume a finite magnetic unit cell, parameterize all spins in this unit cell by two angles each, and minimize the classical energy as function of these angles. In both 2D and 3D, we have included all possible magnetic unit cells with up to 12 sites. In 2D and 3D, this leads to 8 and 21 different unit-cell geometries, respectively (plus symmetry-equivalent ones).
Among the phases known for the 2D HK model in a field, this approach captures all phases apart from the ``zigzag-star'' and ``diluted-star'' phases of Ref.~\onlinecite{janssen2016}, which occur at elevated $[111]$ fields above the zigzag and stripy phases, respectively. In Ref.~\onlinecite{janssen2016}, these phases were found to have a 36-site (zigzag star) and 18-site (diluted star) unit cell, but it is likely that their true structures are incommensurate.\cite{chern2017}

We have also determined the phase diagram of the hyperhoneycomb HK model in a field along the $[001] \propto \mathbf{\hat z}$ direction, for which all in-field ordered states are simply canted versions of the zero-field orders, without any field-induced intermediate phases, in complete quantitative equivalence with the 2D honeycomb-lattice result.\cite{janssen2016}

\subsection{Incommensurate phases of the HK$\boldsymbol{\pm\Gamma}$ model}

In the phase diagram of the HK$\pm\Gamma$ model, a prominent role is taken by incommensurate spiral phases.\cite{lee2015} Hence, an energetic minimization assuming a finite unit cell is inefficient. Instead, we use a combination of the Luttinger-Tisza approach \cite{luttinger1946,kaplan1960} and a single-$\mathbf Q$ ansatz.
The Luttinger-Tisza approach captures both commensurate and incommensurate magnetic states, but cannot discriminate between single-$\mathbf Q$ and multi-$\mathbf Q$ cases. 
Moreover, in situations in which the crystallographic unit cell consists of more than one site, the method not necessarily finds the true classical ground state.
Single-$\mathbf Q$ spiral-type states are most efficiently captured with the ansatz
\begin{equation}
 \mathbf{S}_a(\mathbf{r}) = 
 \left[
 \mathbf{\hat{e}}^x_a\cos(\mathbf{Q}\cdot\mathbf{r})+\mathbf{\hat{e}}^y_a\sin(\mathbf{Q}\cdot\mathbf{r})
 \right] 
 \sin\eta_a 
 + \mathbf{\hat{e}}^z_a \cos\eta_a,
\end{equation}
where $a=1,\ldots, 4$ labels the sublattice and $\mathbf{r}$ the position of the spin. The ordering wavevector $\mathbf{Q}$, the basis vectors $\{\mathbf{\hat{e}}^x_a,\mathbf{\hat{e}}^y_a,\mathbf{\hat{e}}^z_a\}$ and the canting parameters $\eta_a$ are variational parameters.

%%%%%%%%%%%%%%%%%%%%%%%%%%%%%%%%%%%%%%%%%%%%%%%%%%%%%%%%%%%%%%%%%%%%%%%
\begin{table}[t]
\caption{Characterization of ordered phases on the hyperhoneycomb lattice in the HK model in a $[111]$ magnetic field and the HK$\Gamma$ model at zero field.
We have adopted the nomenclatures introduced in Refs.~\onlinecite{janssen2016, lee2015}.
The position of the high-symmetry points in the $\{\mathbf{a^*}, \mathbf{b^*}, \mathbf{c^*}\}$ basis are:
$\boldsymbol{\Gamma} = (0,0,0)$,
$\mathbf{Y} = (0,-\frac12,0)$,
$\mathbf{E} = (-\frac{1}{3},0,0)$.
Except for the 3D spiral, AF$_{abc}$, and $\overline{\text{SP}}_{b^\pm}$ phases, all states are quasi-2D, i.e., they can be mapped onto equivalent states on the honeycomb lattice.}
\begin{tabular*}{\linewidth}{@{\extracolsep{\fill} } lll}
\hline\hline
Phase & $\mathbf Q$ & Comments\\ \hline
skew-zigzag (SZ) & $\boldsymbol\Gamma$ or $\mathbf Y$ & $\mathbf S_i \parallel \mathbf{\hat x}$ or $\mathbf S_i \parallel \mathbf{\hat y}$ or $\mathbf S_i \parallel \mathbf{\hat z}$ (degenerate) \\
AF star & $\boldsymbol\Gamma$ and $\mathbf Y$ & multi-$\mathbf Q$\\
AF vortex & $\mathbf E$ & \\
3D spiral & $(0,\frac13,0)$ & genuine 3D\\
SZ$_{x/y}$ & $\mathbf Y$ & zigzag in $\mathbf S_i \cdot \mathbf{\hat x}$ or $\mathbf S_i \cdot \mathbf{\hat y}$ (degenerate) \\ 
SZ$_{b}$ & $\boldsymbol\Gamma$ & zigzag in $\mathbf S_i \cdot \mathbf{\hat b}$\\
skew-stripy (SS) & $\boldsymbol\Gamma$ or $\mathbf Y$ & $\mathbf S_i \parallel \mathbf{\hat x}$ or $\mathbf S_i \parallel \mathbf{\hat y}$ or $\mathbf S_i \parallel \mathbf{\hat z}$ (degenerate) \\
FM star & $\boldsymbol\Gamma$ and $\mathbf Y$ & multi-$\mathbf Q$\\
vortex & $\mathbf E$ & \\
SS$_{x/y}$ & $\mathbf Y$ & stripy in $\mathbf S_i \cdot \mathbf{\hat x}$ or $\mathbf S_i \cdot \mathbf{\hat y}$ (degenerate) \\ 
N\'eel & $\boldsymbol\Gamma$ & \\
AF$_a$ & $\boldsymbol\Gamma$ & N\'eel in $\mathbf S_i \cdot \mathbf{\hat a}$, stripy in $\mathbf S_i \cdot \mathbf{\hat b}$\\
AF$_{abc}$ & $(\frac14, \frac14, \frac14)$ & genuine 3D\\
ferromagnet (FM)& $\boldsymbol\Gamma$ & \\
polarized & $\boldsymbol\Gamma$ & $\mathbf S_i \parallel \mathbf h$ \\
FM$_c$ & $\boldsymbol\Gamma$ & $\mathbf S_i \parallel \mathbf{\hat c}$\\
FM-SZ$_\text{FM}$ & $\boldsymbol\Gamma$ & FM in $\mathbf S_i \cdot \mathbf{\hat b}$, zigzag in $\mathbf S_i \cdot \mathbf{\hat a}$\\
$\overline{\mathrm{SP}}_{a^\pm}$ & $(h,0,0)$ & incommensurate spiral, quasi-2D \\
$\overline{\mathrm{SP}}_{b^\pm}$ & $(0,k,0)$ & incommensurate spiral, genuine 3D\\
\hline
\end{tabular*}
\label{tab:phases}
\end{table}
%%%%%%%%%%%%%%%%%%%%%%%%%%%%%%%%%%%%%%%%%%%%%%%%%%%%%%%%%%%%%%%%%%%%%%%

%%%%%%%%%%%%%%%%%%%%%%%%%%%%%%%%%%%%%%%%%%%%%%%%%%%%%%%%%%%%%%%%%%%%%%%
\begin{figure*}
\includegraphics[scale=.92]{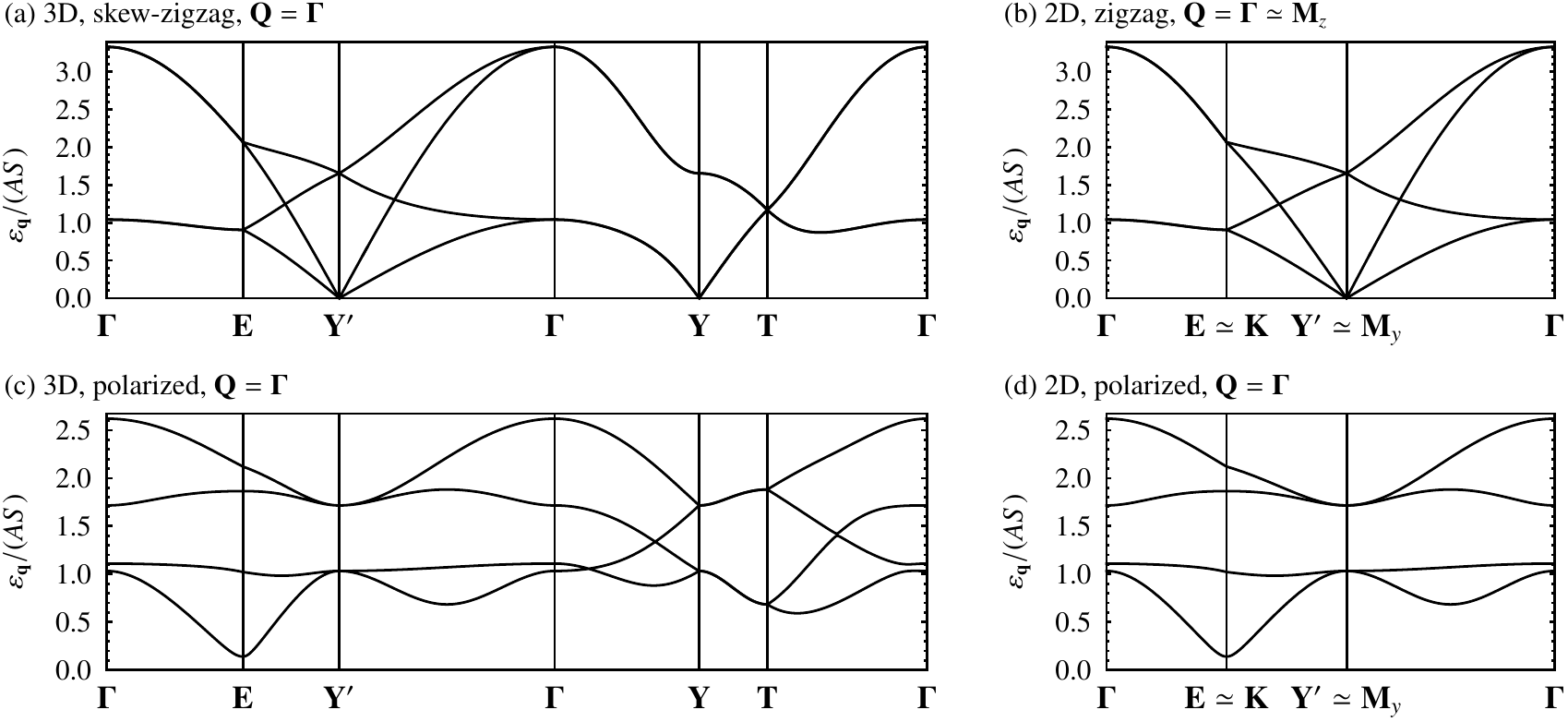}
\caption{Top panels: Magnon spectrum along high-symmetry points in the zigzag phase of the HK model for $\varphi = 0.62\pi$ (in the conventions of Fig.~2 of the main paper) and $h=0$ on the (a) hyperhoneycomb and (b) honeycomb lattices.
The position of the high-symmetry points in the $\{\mathbf{a^*}, \mathbf{b^*}, \mathbf{c^*}\}$ basis are:
$\boldsymbol{\Gamma} = (0,0,0)$,
$\mathbf{E} = (-\frac{1}{3},0,0)$,
$\mathbf{Y'} = (-\frac{1}{2}, 0, \frac{1}{2})$,
$\mathbf{Y} = (0,-\frac12,0)$,
$\mathbf{T} = (0,-\frac12,-\frac12)$.
Note that the spectrum along the ``quasi-2D'' path $\boldsymbol{\Gamma} \to \mathbf{E} \to \mathbf{Y}' \to \boldsymbol{\Gamma}$ is exactly equal to the honeycomb-lattice case.
Bottom panels: Same as top panels, but in the field-polarized phase for $h/(AS)=2.62$ with field applied in the $[111]$ direction, i.e., slightly above the critical field.
To demonstrate the equivalence with (c), a rectangular four-site magnetic unit cell has been chosen in (d) as well. Within this band backfolding, the corner of the hexagonal Brillouin zone $\mathbf K = (-\frac23,0,0)$ is equivalent to $\mathbf E$ in the rectangular Brillouin zone, and the centers of the edges of the hexagonal Brillouin zone $\mathbf M_y = (-\frac12,0,\frac12)$ and $\mathbf M_z = (0,0,-1)$ are equivalent to $\mathbf{Y}'$ and $\boldsymbol\Gamma$ in the rectangular Brillouin zone.
}
\label{fig:spectra}
\end{figure*}
%%%%%%%%%%%%%%%%%%%%%%%%%%%%%%%%%%%%%%%%%%%%%%%%%%%%%%%%%%%%%%%%%%%%%%%

For the HK$\pm\Gamma$ model, both approaches delivered identical results in most of the phase diagram, with the exception of a region (marked by white lines in Fig.~3 of the main paper) where the Luttinger-Tisza solution violated the spin-length constraint. Consequently, we have used the results of the single-$\mathbf Q$ ansatz in these cases.

\subsection{Summary: Quasi-2D and genuine 3D phases}

The various phases that have been obtained in the HK model in a $[111]$ magnetic field and in the HK$\Gamma$ model at zero field are characterized in Table~\ref{tab:phases}.

In the HK model in an applied field, there is a single field-induced phase that is of genuine 3D character, denoted as ``3D spiral'' in Fig.~2(a,c) of the main paper and in Table~\ref{tab:phases}. 
This phase consists of period-three spirals along the $\mathbf b$ axis, with parameter-dependent angles between spins separated by $\mathbf b$ forming an $AAB$ pattern.
The ordering wavevector is consequently given by $\mathbf Q = \frac13 \mathbf{b}^*$, which is not equivalent to any point in the $ac$ plane. The 3D spiral state can therefore not be mapped onto the 2D honeycomb lattice.

In the HK$\pm\Gamma$ model, we find two genuine 3D phases, labeled as AF$_{abc}$ and $\overline{\mathrm{SP}}_{b^\pm}$. The former is a commensurate phase with a period-two modulation along the vector $\mathbf a+\mathbf b+\mathbf c$. The latter consists of incommensurate spirals formed along the $\mathbf b$ direction. All other phases are quasi-2D and are found, with the same energetics, also in the honeycomb-lattice HK$\pm\Gamma$ model.

%%%%%%%%%%%%%%%%%%%%%%%%%%%%%%%%%%%%%%%%%%%%%%%%%%%%%%%%%%%%%%%%%%%%%%%
%%%%%%%%%%%%%%%%%%%%%%%%%%%%%%%%%%%%%%%%%%%%%%%%%%%%%%%%%%%%%%%%%%%%%%%
%%%%%%%%%%%%%%%%%%%%%%%%%%%%%%%%%%%%%%%%%%%%%%%%%%%%%%%%%%%%%%%%%%%%%%%

\section{Magnon spectra and quantum corrections}

Here we discuss corrections to the classical picture, which formally corresponds to the limit of infinite spin magnitude, $S\to\infty$. At $T=0$, quantum corrections from $S<\infty$ are different in 3D and 2D and will hence invalidate the exact 3D--2D mapping. Nevertheless, the mapping continues to apply on a qualitative level for most regions of the phase diagrams.

Spin-wave theory is the canonical method to determine excitation spectra and quantum corrections for magnetically ordered states in spin models. Spin-wave theory can be organized in a systematic expansion in $1/S$. Here we determine the leading-order spin-wave spectra which then can be used to calculate thermodynamic observables to next-to-leading order, most importantly the order parameter and the uniform magnetization.

We employ the Holstein-Primakoff representation with distinct boson operators on the four (two) sublattices on the hyperhoneycomb (honeycomb) lattice. The resulting Bogoliubov problem is diagonalized numerically. As the methodology is standard, we refer the reader to the review in Ref.~\onlinecite{janssen2019} for details.

\subsection{3D--2D mapping of spectra}

Inspecting the leading-order spin-wave spectra reveals a remarkable fact: For 3D momenta that are ``quasi-2D'', i.e., that thave a vanishing component along $\mathbf{b}^*$, the mode energies in the 3D case are exactly equal to the mode energies of the corresponding 2D model at the respective projected momenta. This is illustrated for the HK model in Fig.~\ref{fig:spectra} for both the zero-field zigzag phase and the high-field phase; to facilitate the comparison, a four-site unit cell has been chosen in the 2D case.
At zero field, pseudo-Goldstone modes are seen at $\mathbf{Y}$ and $\mathbf{Y}'$ wavevectors, which are shifted from the $z$ skew-zigzag ordering wavevector $\mathbf Q = \boldsymbol{\Gamma}$, in full agreement with the situation on the honeycomb lattice.\cite{chaloupka2015} In the high-field phase, all modes are gapped.

The 3D--2D mapping of the magnon spectra can be understood as follows: Given the equivalence of the lattice structures (and interactions) after projection on the $ac$ plane, the Fourier-transformed interactions appearing in the magnon Hamiltonian are equivalent after projection as well, i.e., they are equal for quasi-2D momenta. It is these (bare) interactions that determine the magnon spectra at leading order in $1/S$, hence the 3D--2D mapping is generically valid.

At a given wavevector, the leading-order dynamic spin structure factor depends only on the components of the Bogoliubov transformation matrix at this wavevector.\cite{janssen2019} The 3D--2D equivalence therefore holds for both the static and dynamic structure factors, provided the restriction to the $ac$ plane in the latter case.

The equivalence breaks down at higher orders in $1/S$ because quantum corrections arising from magnon-magnon interactions involve momentum integrals running over the full Brillouin zone in either 3D or 2D and hence are different in both cases. The same applies to quantum corrections to thermodynamic quantities, as shown below.

%%%%%%%%%%%%%%%%%%%%%%%%%%%%%%%%%%%%%%%%%%%%%%%%%%%%%%%%%%%%%%%%%%%%%%%
\begin{figure}[t]
\includegraphics[width=.95\linewidth]{{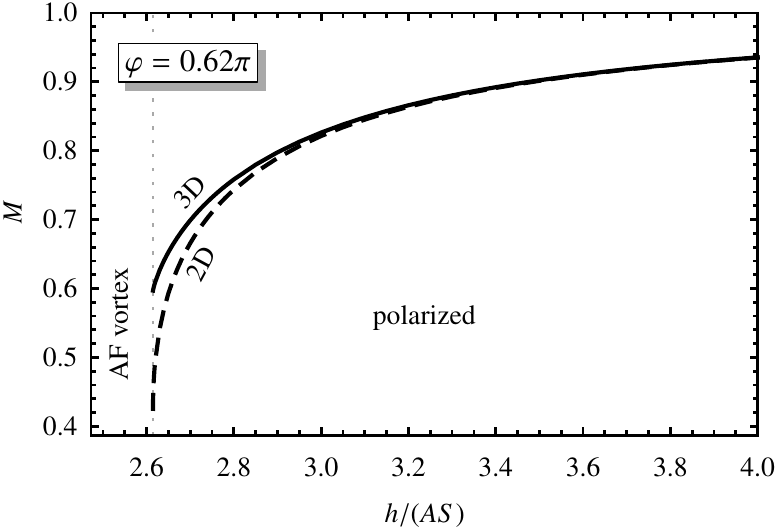}}
\caption{
Magnetization in the field-polarized phase for $\varphi = 0.62\pi$ of the HK model as a function of the magnetic field applied in $[111]$ direction, obtained from linear spin-wave theory for $S=1/2$, comparing the hyperhoneycomb (solid) and honeycomb (dashed) lattices.
}
\label{fig:magpol}
\end{figure}
%%%%%%%%%%%%%%%%%%%%%%%%%%%%%%%%%%%%%%%%%%%%%%%%%%%%%%%%%%%%%%%%%%%%%%%

\subsection{Quantum corrections to magnetization}

The spin-wave results can be used to determine zero-temperature quantum corrections to local expectation values.
We employ this to calculate the magnetization in the high-field phase of the HK model. Here, the magnetization points in field direction and is saturated in the classical limit. Using the Holstein-Primakoff representation, the magnetization~$M$ along the field direction including quantum corrections is
\begin{equation}
 M = 
 S - \frac{1}{N}\sum_\mathbf{q}
 \left\langle 
 a^\dagger_\mathbf{q}a_\mathbf{q}+b^\dagger_\mathbf{q}b_\mathbf{q}+c^\dagger_\mathbf{q}c_\mathbf{q}+d^\dagger_\mathbf{q}d_\mathbf{q}
 \right\rangle
\label{mcorr}
\end{equation}
where $a_\mathbf{q}$, $b_\mathbf{q}$, $c_\mathbf{q}$, and $d_\mathbf{q}$ are the magnon operators on the four sublattices, and $N$ is the number of sites. In next-to-leading order in $1/S$, this can be expressed in terms of the Bogoliubov coefficients obtained from linear spin-wave theory, for details see Refs.~\onlinecite{janssen2016,janssen2019}.
The numerical result, evaluated for $S=1/2$, is shown in Fig.~\ref{fig:magpol} for a particular value of $\varphi$, parameterizing the ratio between Heisenberg and Kitaev couplings according to $J=A\cos\varphi$ and $K=2A\sin\varphi$. The comparison between the 2D honeycomb and 3D hyperhoneycomb results confirms that quantum effects are stronger in lower dimensions, as a result of the larger fraction of low-energy phase space.

We have also calculated the quantum corrections to the order parameter of selected zero-field phases: For collinear states this computation involves only amplitude corrections and is similar to that leading to the uniform magnetization in Eq.~\eqref{mcorr}; for non-collinear states angle corrections would need to be considered as well.
Numerical results, again evaluated for $S=1/2$, are shown for the N\'eel and zigzag phases of the HK model in Fig.~2(d) of the main paper. Also here, quantum effects are stronger in lower dimensions.

These results illustrate that quantum corrections invalidate the 3D--2D mapping for $S < \infty$. However, thermodynamic observables, such as magnetization curves, continue to agree in 3D and 2D on a qualitative level in large regions of the phase space, in particular deep inside magnetically ordered and classical paramagnetic phases.

\section{Implications for observables}

Here, we summarize the implications of the 3D--2D mapping for experimental and theoretical observables. In addition to the total energy and the magnetization for $S \to \infty$, momentum-resolved observables enjoy the 3D--2D equivalence for the 3D wavevectors with zero component along $\mathbf{b}^*$. This includes the static and dynamic spin structure factors, as measured in neutron scattering. In contrast, other observables that are momentum-integrated, such as specific heat or magnon thermal conductivity, are different in 3D and 2D due to the different available phase spaces.
The implications of the 3D--2D mapping for various observables to leading order in $1/S$ are compiled in Table~\ref{tab:observables}.

%%%%%%%%%%%%%%%%%%%%%%%%%%%%%%%%%%%%%%%%%%%%%%%%%%%%%%%%%%%%%%%%%%%%%%%
\begin{table}[!b]
\caption{Implications of the 3D--2D mapping for different observables to leading order in $1/S$, assuming that the ground state is quasi-2D.}
\begin{tabular*}{\linewidth}{@{\extracolsep{\fill} }ll}
\hline\hline
Observable & 3D--2D equivalence \\
\hline
Ground-state energy $\varepsilon_0$ & fully applies \\
Magnon spectrum $\varepsilon_{\mathbf q}$ & applies for $\mathbf q \in ac$ plane \\
Bragg-peak positions & fully applies \\
Dynamic structure factor $\mathcal S(\mathbf q,\omega)$ & applies for $\mathbf q \in ac$ plane \\
Order parameter $M_\text{stagg}$ & fully applies \\
Uniform magnetization $M$ & fully applies \\
Specific heat $C_\text{mag}$ & no exact equivalence \\
Magnon thermal conductivity $\kappa$ & no exact equivalence \\
\hline \hline
\end{tabular*}
\label{tab:observables}
\end{table}
%%%%%%%%%%%%%%%%%%%%%%%%%%%%%%%%%%%%%%%%%%%%%%%%%%%%%%%%%%%%%%%%%%%%%%%

%%%%%%%%%%%%%%%%%%%%%%%%%%%%%%%%%%%%%%%%%%%%%%%%%%%%%%%%%%%%%%%%%%%%%%%